\documentclass[12pt]{article}

\usepackage{amsmath}
\usepackage{amssymb}
\usepackage{amsthm}

\usepackage{upref}

\newtheorem{thm}{Theorem}[section]

\def\d{{\delta}}

\def\a{{\alpha}}
\def\e{{\varepsilon}}
\def\l{{\lambda}}

\def\F{{\mathbb F}}

\def\Q{{\bf Q}}
\def\R{{\mathbb R}}
\def\B{{\mathbb B}}
\def\X{{\mathbb X}}
\def\M{{\rm Min}}
\def\hh{{H}}

\title{\bf The Sturm-Liouville eigenvalue problem\\ 
and  NP-complete problems \\
in the quantum setting with queries}
\author{A. Papageorgiou$^1$ and H. Wo\'zniakowski$^2$ \\
{\small $^{1,2}$Department of Computer Science, Columbia University, 
New York, USA} \\
{\small $^2$Institute of Applied Mathematics and Mechanics, 
University of Warsaw, Poland}}

\date{\today}
\begin{document}

\setcounter{page}{1}
\maketitle

\begin{abstract} 
We show how a number of NP-complete as well as NP-hard problems 
can be reduced to the Sturm-Liouville eigenvalue problem 
in the quantum setting with queries.
We consider {\it power} queries which are derived from the propagator
of a system evolving with a Hamiltonian obtained from the discretization of 
the Sturm-Liouville operator.
We show that the number of power queries as well the number of qubits needed 
to solve the problems studied in this paper 
is a low degree polynomial.
The implementation of power queries 
by a polynomial number of elementary quantum gates is an open issue.
If this problem is solved positively 
for the power queries used for the Sturm-Liouville eigenvalue problem 
then a quantum computer would be a very powerful
computation device allowing us to solve
NP-complete problems in polynomial time.
\end{abstract}

\section{Introduction}

An important question in quantum computing is whether NP-complete
problems can be solved in polynomial time, see 
\cite{bennet,bernstein} and papers cited there. 
We address this question by studying the quantum setting with queries.
We consider two types of queries: {\it bit} and {\it power} queries, 
see \cite{nielsen} for general information about queries and 
quantum computation. Here we only mention that bit queries are used in
Grover's search algorithm \cite{grover}.
They allow us to obtain the values of 
Boolean functions \cite{bennet}, and the approximate values of
real functions \cite{heinrich}. Moreover,  we know 
that bit queries cannot be used to solve NP-complete problems in polynomial 
time \cite{bennet}.
Power queries are used in the well-known phase estimation
algorithm, see \cite{nielsen}, 
which plays a central role in Shor's
factorization algorithm \cite{shor}. In a recent paper \cite{PW1} we
dealt with power queries in the study of the quantum complexity of 
the Sturm-Liouville eigenvalue problem. 

In this paper, we show how to reduce NP-complete problems to 
the Sturm-Liouville eigenvalue problem whose complexity in the classical
and quantum settings has been studied in \cite{PW1}. 
Obviously, it would be enough to show this reduction for one
NP-complete problem. We choose to present this reduction for several
problems to show how the number of power queries and qubits 
depends on the particular NP-complete problem.
In particular, that is why we consider
satisfiability and the traveling
salesman problem, as well as their NP-hard versions.  
The reductions presented in this paper can be summarized in the 
following diagram.
\begin{eqnarray*}
\mbox{{\bf SAT}}\ \implies && \mbox{{\bf BOOL}}\ \implies\ \mbox{{\bf INT}}\ \implies\ \mbox{{\bf SLE}}\\
\mbox{{\bf TSP}}\ \implies\ \mbox{{\bf MIN}}\ \implies && \mbox{{\bf BOOL}}\ \implies\ \mbox{{\bf INT}}\ \implies\ \mbox{{\bf SLE}}\\
\mbox{{\bf GRO}}\ \implies && \mbox{{\bf BOOL}}\ \implies\ \mbox{{\bf INT}}\ \implies\ \mbox{{\bf SLE}}
\end{eqnarray*}
Here, {\bf SAT} stands for the satisfiability problem, {\bf TSP} for the 
traveling salesman problem, {\bf MIN} for
the minimization problem of choosing the smallest number out of $N$ real 
numbers, {\bf GRO} for Grover's problem,
{\bf BOOL} for the Boolean mean problem, {\bf INT} for the integration problem,
and finally {\bf SLE} for the Sturm-Liouville eigenvalue problem.

These reductions mean, in particular, that the satisfiability problem is reduced to the Boolean mean problem for a specific Boolean function
which is reduced to the integration problem for a specific integrand, which in turn is reduced to the Sturm-Liouville eigenvalue problem
for a specific function, and finally the last problem is solved by the quantum algorithm using power queries.  

The Sturm-Liouville
problem is defined in the next section. For the moment we mention that we
want to approximate the smallest eigenvalue of a specific 
differential operator, and this smallest eigenvalue
is given in a variational form as the minimum of specific integrals. 
We use a formula relating the Sturm-Liouville
eigenvalue problem to a weighted integration problem, see \cite{PW1}.
Many computational problems including the discrete problems mentioned above  
can be recasted as this weighted integration problem. Thus,
we can solve them  using the algorithms of
\cite{PW1} for solving the Sturm-Liouville eigenvalue problem.
These algorithms use of order $\e^{-1/3}$ bit queries or
$\log\,\e^{-1}$ power queries and compute  an $\e$-approximation 
of the smallest eigenvalue with probability $\tfrac34$.
The bounds on bit and power queries are sharp up to multiplicative
constants, see \cite{Bessen,PW1}. Hence, exponentially fewer
power queries than bit queries are needed 
to solve the Sturm-Liouville eigenvalue problem. 
As we shall see, the same is true for the problems studied 
in this paper.

In the quantum setting with bit queries, we do not obtain 
surprising results.  
The polynomial number of bit queries, $\e^{-1/3}$, 
implies that the solution of NP-complete problems by 
modifications of the algorithm for the Sturm-Liouville eigenvalue 
problem will require exponentially many queries in terms of the NP 
problem size.

The situation is quite different if we consider
power queries. The logarithmic number of power queries, $\log\,\e^{-1}$, 
implies that NP-complete problems can be solved by 
modifications of the algorithm for the Sturm-Liouville eigenvalue
problem and the number of power queries is polynomial in the problem size. 

More specifically, the satisfiability problem for Boolean functions 
with $n$ variables can be solved with probability $1-\d$  using of 
order $n\,\log\,\d^{-1}$ power queries and $n$ qubits.
Furthermore, a truth assignment to a non-zero Boolean function with
$n$ variables can be computed with probability $1-\d$ using of order
$n^2(\log\,\d^{-1}+\log\,n)$ power queries and $n$ qubits.

The traveling salesman problem with $m$ cities can be solved with
probability $1-\d$ using of order 
$$
m\log\,m(\log\,\d^{-1}+\log\,m+\log\,
d_{\max})(\log\,m+\log\,d_{\max})
$$
power queries and $m\log\,m$ qubits, 
where $d_{\max}$ denotes the maximal distance between cities. 
Furthermore, an optimal route for the traveling salesman problem 
can be computed with probability $1-\d$ using of order
$$
m^2\log^2m(\log\,\d^{-1}+\log\,m)+
m\log\,m(\log\,\d^{-1}+\log\,m+\log\,
d_{\max})(\log\,m+\log\,d_{\max})
$$
power queries and $m\log\,m$ qubits. 
 
Finally, Grover's problem for Boolean functions with $n$ variables 
can be solved with  probability $1-\d$ using of order 
$n\log\,\d^{-1}$ power queries and $n$ qubits. 
It is well known that
Grover's problem requires of order $2^{n/2}$ bit queries. Therefore,
it is evident that exponentially fewer power than bit queries
are required for this problem.

We stress that we only show how many power queries
are needed to solve a particular problem. 
We use power queries which are of the form
controlled-~$W^{p_j}$ 
for a $k\times k$ unitary
matrix $W$ and some exponents $p_j$. 
In our case, the matrix $W$ is given by
\begin{equation}\label{1}
W\,=\,\exp\left(\tfrac12\,\mathrm{i}\,M_q\right)\qquad \mbox{with}\ \ 
\mathrm{i}=\sqrt{-1}.
\end{equation}
Here, the matrix $M_q$ has a particularly simple form since it is
a $k\times k$  real symmetric tridiagonal matrix, 
$$
M_q\,=\,(k+1)^2\,
\left[
\begin{array}{ccccc}
2 & -1 & & & \\
-1 & 2 & -1 & &\\
   &  \ddots & \ddots & \ddots & \\
   &         & -1 & 2 & -1 \\
   &         &    & -1 & 2
\end{array}
\right] + \left[
\begin{array}{ccccc}
q(\tfrac1{k+1}) & & & & \\
& q(\tfrac2{k+1}) & & &  \\
& & \ddots  & &  \\
& & & q(\tfrac{k-1}{k+1}) &  \\
& & &  & q(\tfrac{k}{k+1})
\end{array} \right],
$$
with a function 
$q:[0,1]\to[0,1]$ 
that is two times continuously differentiable and bounded by one up to the
second derivative.  This matrix corresponds to 
a classical approximation of the Sturm-Liouville operator.

Contrary to the situation in Shor's algorithm where powers of a unitary 
operator can be implemented efficiently, the quantum implementation of 
the power queries for the $k\times k$ matrix $W$ of 
the form (\ref{1}) by a number of known elementary quantum gates 
which is polylog in $k$ is an open issue.
If it turns out that the implementation cost of such power queries  
is disproportionally large compared to that of bit queries,
then the positive results on the number of power queries 
are of only theoretical interest. If, on the other hand, 
power queries for $W$ of the form (\ref{1}) and  a function $q$ satisfying 
conditions that we discuss later in this paper, can be implemented
efficiently, i.e., at cost which is polylog in $k$, 
then we will have a very powerful computational device 
allowing us to solve NP-complete problems in polynomial time.

\section{Sturm-Liouville and Integration}
We briefly recall the problem and some of the results from \cite{PW1}.
We consider the following class of functions
$$
\Q\,=\,\left\{\,q:[0,1]\to [0,1]\ \bigg|\ \ q\in C^2([0,1])\ \
\mbox{and}\ \ 
\max_{i=0,1,2}\ \|q^{(i)}\|_{\infty}\,\le 1\,\right\},
$$
where $C^2([0,1])$ stands for the class of twice
continuously differentiable functions, and 
$\|q\|_{\infty}=\max_{x\in [0,1]}|q(x)|$.

We studied the approximate computation of the Sturm-Liouville smallest
eigenvalue~$\l(q)$ which is defined in the variational form  by 
\begin{equation}\label{var}
\l(q)\,=\,\min_{0\ne u\in H_0^1} \frac{\int_0^1\left[ (u^\prime(x))^2 + q(x)
u^2(x)\right] \, dx}
{\int_0^1 u^2(x)\, dx}\, ,
\end{equation}
where $H_0^1$ is the Sobolev space of absolutely continuous functions
for which $u^\prime\, \in L_2([0,1])$ and $u(0)=u(1)=0$, 
see \cite{babuska,courant,strang}. Combining results from
from \cite{courant,keller,titschmarsh} we have 
the formula that relates the Sturm-Liouville smallest eigenvalue 
problem to integration,
\begin{equation}\label{444}
\int_0^1\left(q(x)-\tfrac12\right)\sin^2(\pi x)\, dx\,=\,
\tfrac12\left(\l(q)-\pi^2-\tfrac12\right)
\,+\,O\left(\|q-\tfrac12\|_{\infty}^2\right).
\end{equation}

We analyzed the quantum setting with bit and power
queries in \cite{PW1}. For bit queries, we showed that $\l(q)$ 
can be computed with error $\eta$ and probability $\tfrac34$ using 
$\Theta(\eta^{-1/3})$ bit queries, and this bound is sharp
modulo a multiplicative constant. It is easy to check that from this
result follows that NP-complete problems of size $n$
can be solved with an exponential in $n$ number of bit queries. 
Therefore, from now on, we restrict
ourselves to the quantum setting with power queries for matrices $W$ of
the form (\ref{1}). 

In \cite{PW1} we presented a quantum algorithm $\phi$ based
on phase estimation applied to the
discretized matrix of the Sturm-Liouville problem. The initial 
state was an approximate eigenvector, as proposed by
Abrams and Lloyd \cite{abrams}, and computed by the algorithm of
the Jaksch and Papageorgiou \cite{jaksch}.
The algorithm $\phi$ computes 
$\l(q,\eta)$ such that
$$
|\l(q)-\l(q,\eta)|\,\le\,\eta\qquad\mbox{with probability}\ \ \tfrac34
\quad \forall\,q\in\Q
$$
using of order $\log\eta^{-1}$ power queries, $\log^2\eta^{-1}$
additional quantum operations, $O(1)$ function values of $q$ 
and classical operations, as well as $\log\eta^{-1}$ qubits. 

It is well known that we can increase the probability of success 
to, say, $1-\delta$ by repeating 
the algorithm $\phi$ of order $\log \delta^{-1}$ times and then
taking the median as the final approximation. The algorithm $\phi$ 
with repetitions computes $\l(q,\eta,\d)$ which
is an $\eta$-approximation of $\l(q)$ with probability $1-\d$, i.e., 
\begin{equation}\label{alg}
|\l(q)-\l(q,\eta,\d)|\,\le\,\eta\qquad\mbox{with probability}\ \ 1-\d 
\quad \forall\,q\in\Q.
\end{equation}
The resulting algorithm with repetitions $\phi$ uses of order
\begin{itemize}
\item 
$\log\d^{-1}\,\log\eta^{-1}$ power queries,
\item 
$\log\d^{-1}\,\log^2\eta^{-1}$ additional quantum operations,
\item
$O(1)$ function values of $q$ and classical operations, and 
\item $\log\eta^{-1}$  qubits.
\end{itemize}

In the next sections we show how to modify the algorithm $\phi$ 
to solve a number of continuous and discrete problems.
In what follows, we restrict ourselves and mention only the number
of power queries and the number of qubits of these modifications
because they are the most important characteristic of the cost of a quantum
algorithm. The rest of the cost characteristics
can be easily derived from the corresponding components of the 
cost of the algorithm $\phi$ with repetitions.

We start with integration. 
Knowing how to approximate $\l(q)$ we can approximate the integral 
in (\ref{444}) modulo the second term which is of order 
$\|q-\frac12\|^2_{\infty}$. We provide the details in the next section.

\section{Integration}\label{inte}
Consider the (weighted) integration problem
$$
I(f)\,:=\,\int_0^1f(x)\,\sin^2(\pi x)\,dx
$$
for functions $f$ from the class 
$$
\F_M\,=\,\big\{f\in C^2([0,1]):\ \max_{i=0,1,2}
\|f^{(i)}\|_{\infty}\,\le M\,\big\}.
$$
Here $M$ is a positive number. We want to compute an
$\e$-approximation of $I(f)$ with probability $1-\d$ on a quantum
computer with power queries. 
Since $|I(f)|\le M$ we assume that $\e<M$ since otherwise
$0$ is an $\e$-approximation, and the problem is trivial. 
Without loss of generality, we also assume that $\e<1$.

Observe that $f\in \F_M$ implies that the function 
$$
q_{f,c}(x)\,=\,\tfrac12\,+\,cf(x)\qquad \forall\,x\in[0,1],
$$
belongs to $\Q$, defined in the previous section, for $c\in(0,(2M)^{-1}]$. 
In this case, the 
formula (\ref{444}) states
$$
I(f)\,=\,\frac1{2c}\left(
\l(q_{f,c})-\pi^2-\tfrac12\right)\,+\,O(cM^2)\qquad\forall\,f\in \F_M.
$$

Define 
$$
c\,=\,\frac{\e}{M^2\,\log\,\e^{-1}}\quad\mbox{and}\quad
\eta\,=\,c\,\e\,=\,\frac{\e^2}{M^2\log\,\e^{-1}}.
$$ 
Let $\l(q_{f,c},\eta,\d)$ be an $\eta$-approximation of $\l(q_{f,c})$ with
probability $1-\d$ computed by the algorithm $\phi$ with
repetitions of the previous section. 
Knowing $\l(q_{f,c},\eta,\d)$ we compute on a classical computer
$$
A^{{\rm Int}}(f,\e,\d)\,=\,
\frac1{2c}\left(\l(q_{f,c},\eta,\d)-\pi^2-\tfrac12\right).
$$
Then
$$
|I(f)-A^{{\rm Int}}(f,\e,\d)|\,\le\,\frac1{2c}\,
\eta\,+\,O(cM^2)\,=\,\tfrac12\,\e(1+o(1)) \ \ \
\mbox{with probability}\ \ 1-\d.
$$
Hence, for small $\e$, $A^{{\rm Int}}(f,\e,\d)$ is an
$\e$-approximation of $I(f)$ with probability $1-\d$. 
In this way we can solve the integration
problem and we summarize this result in the following theorem.
\begin{thm}\label{3.1}
We compute an $\e$-approximation with probability 
$1-\delta$ for the integration problem for the class $\F_M$ 
by the quantum algorithm $A^{{\rm Int}}$ using of order
\begin{itemize}
\item
$\log\d^{-1}\,\left(\log\,M\,+\,\log\,\e^{-1}\right)$
 power queries and
\item 
$\log\,M\,+\,\log\,\e^{-1}\quad\mbox{qubits}$.
\end{itemize}
\end{thm}

\section{Preliminaries}\label{pre}
We now  present some preliminaries that will be used as technical 
tools to translate the integration problem of the previous section to 
the NP-complete and NP-hard problems discussed in this paper.

Take a function $h\in C^2([0,1])$  with $h^{(i)}(0)=h^{(i)}(1)=0$ 
for $i=0,1,2$, and for which the integral $\int_0^1h(x)\,dx$ is positive. 
Examples of such functions include $h(x)=(x(1-x))^{\a}$
with $\a>2$ or $h(x)=x^3(1-x)^3g(x)$ for a positive $g\in
C^2([0,1])$. We extend
the domain of $h$ by defining $\hh(x)=h(x)$ for $x\in[0,1]$ and
$\hh(x)=0$ otherwise. Due to the boundary conditions imposed on $h$, we
have $\hh\in C^2(\R)$.   

For a positive (large) integer $N$, we subdivide the interval
$[\tfrac14,\tfrac34]$ by introducing the points
$$
x_j\,=\,\frac14\,+\,\frac12\,\frac{j}N\qquad\mbox{for}\  \ j=0,1,\dots,N.
$$
For $j=0,1,\dots,N-1$, we define the functions
$$
h_j(x)\,=\,\frac1{4N^2}\,\hh\left(2N(x-x_j)\right)\qquad
\mbox{for}\ \  x\in [0,1].
$$
Observe that $h_j$ vanishes outside the open interval $(x_j,x_{j+1})$,
and $\|h_j\|_{\infty}=\|h\|_{\infty}/(4N^2)$,
$\|h_j^{\prime}\|_{\infty}=\|h^{\prime}\|_{\infty}/(2N)$, and
$\|h^{\prime\prime}_j\|_{\infty}=\|h^{\prime\prime}\|_{\infty}$.
Hence, if we set 
$$
M\,:=\,
\max\bigg(\frac{\|h\|_{\infty}}{4N^2},\frac{\|h^{\prime}\|_{\infty}}{2N}, 
\|h^{\prime\prime}\|_{\infty}\bigg),
$$
which is equal to $\|h^{\prime\prime}\|_{\infty}$ 
for large $N$, then 
$$
h_j\,\in \F_M\qquad \mbox{for}\ \ j=0,1,\dots,N-1.
$$
Observe finally that
\begin{equation}\label{hj}
\int_0^1h_j(x)\,dx\,=\,\int_{x_j}^{x_{j+1}}h_j(x)\,dx\,=\,
\frac1{8N^3}\,\mbox{Int}(h)
\qquad
\mbox{for}\ \  j=0,1,\dots,N-1,
\end{equation}
where 
$$
\mbox{Int}(h)\,:=\,\int_0^1h(x)\,dx.
$$

\section{Boolean mean}\label{Bms}
Consider the class $\B_n$ of all Boolean functions of $n$ variables
mapping $\{0,1\}^n$ into $\{0,1\}$. We can equivalently assume that
the domain of such Boolean functions is $\{0,1,\dots,N-1\}$ with
$N=2^n$. Hence, $\B\in \B_n$ means that
$$
\B:\{0,1,\dots,N-1\}\,\to\,\{0,1\}.
$$
We want to approximate the mean
$$
\mbox{S}_N(\B)\,=\,\frac1N\sum_{j=0}^{N-1}\B(j)
$$
by a quantum algorithm with power queries.
  
We now show how this problem can be reduced to the integration problem
of Section \ref{inte}. Using the notation 
of Section \ref{pre}, we define the function $f_{\B}:[0,1]\to\R$ by
\begin{equation*}
f_{\B}(x)\,=\,
\begin{cases}
h_j(x)\,\B(j)\,\left(2\sin^2(\pi x)\right)^{-1} 
&\text{if}\; x\in[x_j,x_{j+1}] \ \text{with} \ j=0,1,\dots,N-1,\\
0 &\text{if} \; x\in[0,\tfrac14]\cup[\tfrac34,1].
\end{cases}
\end{equation*}

Observe that due to the fact that $h_j$ vanishes up to the second
derivatives at $x_j$ and $x_{j+1}$, and the fact that $2\sin^2(\pi
x)\ge1$ for $x\in[\tfrac14,\tfrac34]$, we conclude that $f_{\B}\in
C^2([0,1])$. Furthermore
\begin{eqnarray*}
\|f_{\B}\|_{\infty}&\le&\|h\|_{\infty}/(4N^2),\\
\|f^{\prime}_{\B}\|_{\infty}&\le&\|h^{\prime}\|_{\infty}/(2N)(1+O(N^{-1})),
 and\\
\|f^{\prime\prime}_{\B}\|_{\infty}&\le&\|h^{\prime\prime}
\|_{\infty}(1+O(N^{-1})).
\end{eqnarray*}
Hence, if we set $M=\|h^{\prime\prime}\|_{\infty}(1+O(N^{-1}))$ then
$f_{\B}\in \F_M$. Observe that, due to (\ref{hj}), the integration problem 
for $f_{\B}$ takes now the form
$$
I(f_{\B})\,=\,\sum_{j=0}^{N-1}\tfrac12\B(j)\int_{x_j}^{x_{j+1}}h_j(x)\,dx\,=\,
\frac{\mbox{Int}(h)}{16N^2}\,\frac1N\sum_{j=0}^{N-1}\B(j)\,=\,
\frac{\mbox{Int}(h)}{16N^2}\,\mbox{S}_N(\B).
$$

Let $\eta=\mbox{Int}(h)\e/(16N^2)$. We now use the quantum algorithm
$A^{{\rm Int}}(f_{\B},\eta,\d)$ from Section~\ref{inte} which,
for small $\e$ or large $N$, computes an  
$\eta$-approximation of $I(f_{\B})$ with probability $1-\d$. 
Knowing $A(f_{\B},\eta,\d)$ we compute on a classical computer 
$$
A^{{\rm Bool}}_n(\B,\e,\d)\,=\,\frac{16N^2}{{\rm Int}(h)}\,A^{{\rm Int}}
(f_{\B},\eta,\d).
$$
Then
$$
|\mbox{S}_N(\B)-A^{{\rm Bool}}_n(\B,\e,\d)|\,=\,\frac{16N^2}{{\rm Int}(h)}\,
|I(f_{\B})-A^{{\rm Int}}(f_{\B},\eta,\d)|\,\le
\frac{16N^2}{{\rm Int}(h)}\eta\,=\,\e, 
$$
and this holds with probability $1-\d$.
We summarize this result in the following theorem.
\begin{thm}
We compute an $\e$-approximation with probability 
$1-\delta$ for the Boolean mean problem for the class $\B_n$ 
by the quantum algorithm $A^{{\rm Bool}}_n$ using of order
\begin{itemize}
\item 
$\log\d^{-1}\,\left(\log\,\e^{-1}\,+\,n\right)$ power queries and
\item
$\log\,\e^{-1}\,+\,n$ qubits.
\end{itemize}
\end{thm}
\vskip 1pc
It is known that the amplitude amplification
algorithm of Brassard, H{\o}yer, Mosca and Tapp \cite{brassard} 
computes an $\e$-approximation of $\mbox{S}_N(\B)$ with probability
$8/\pi^2=0.81\dots$ using of order $\min(N,\e^{-1})$ bit queries. 
Furthermore, this number of queries is order-minimal as proven by
Nayak and Wu \cite{nayak}. 
We stress that the basic part of the quantum algorithm
$A^{{\rm Bool}}_n$ is the phase estimation algorithm which uses
power queries. For $\e\ge N^{-1}$ this algorithm uses
of order $\log\,\e^{-1}$ power queries 
to compute an $\e$-approximation with
probability $8/\pi^2$. Hence, 
the nunber of queries has an exponential improvement in 
its the dependence on $\e^{-1}$.
 
The amplitude amplification algorithm of \cite{brassard} has been used as a
basic tool for solving many continuous problems such as 
real mean, multivariate integration, path integration and
multivariate approximation in the quantum setting. These problems have been 
defined over many classical spaces such as $L_p$, Sobolev and Korobov
spaces. For a number of these continuous
problems, the bit query complexity and the quantum speedups 
over the worst case and randomized settings have been established 
based on the optimality of the quantum summation algorithm, see 
\cite{heinrich,H03,H04a,H04b,N01,NSW,TW02}. 
The use of power queries yields an exponential improvement in the
number of queries. This can be achieved simply by
using the exponentially better power query bound for the Boolean mean.

\section{Satisfiability}\label{sat}
The satisfiability problem, SAT for short, is a well known NP-complete
problem in the Turing machine model of computation \cite{garey}. 
This means that all NP-complete problems can be reduced
to SAT in polynomial time, and if the conjecture P$\not=$NP is true
then there are no algorithms solving SAT in polynomial time
with respect to the length of the Boolean function expressed in 
conjunctive normal form.
SAT can be stated as a decision or as a computational problem
and we  deal with both of them in this section. 

As in the previous section, consider the class $\B_n$ of all Boolean
functions of $n$ variables with the domain $\{0,1,\dots,N-1\}$, where $N=2^n$. 
The two variants of SAT problem are defined as
\begin{itemize}
\item\ {\bf $\mbox{SAT}_1:$}\ for $\B\in \B_n$
given in the conjunctive normal form,
verify if there exists
an index $j$ such that $\B(j)=1$.
\item\ {\bf $\mbox{SAT}_2:$}\ 
for a non-zero $\B\in \B_n$ given in the conjunctive normal form,
compute an index $j$ such that $\B(j)=1$.
\end{itemize}

We now show that either problem can be solved with
probability $1-\d$ by using a number of power queries
which is polynomial in $n$ and $\log\,\d^{-1}$.

We begin with $\mbox{SAT}_1$ and use the notation of Section
\ref{Bms}. Observe that for any $\B\in
\B_n$, the mean $\mbox{S}_N(\B)$ is a multiple of $N^{-1}$, 
i.e., $\mbox{S}_N(\B)=k/N$
for some $k\in\{0,1,\dots,N\}$. If we have a real number $x$ such that
$|\mbox{S}_N(\B)-x|<\tfrac1{2N}$ then $|k-N\,x|<\tfrac12$ which implies that 
$$
k\,=\,\left\lfloor N\,x+\tfrac12\right\rfloor \quad
\mbox{and}\quad \mbox{S}_N(\B)\,=\,
\frac{\left\lfloor N\,x+\tfrac12\right\rfloor}{N}.
$$
Obviously, $k>0$ iff there exists an index $j$ for which $\B(j)=1$. 

For $\eta<\tfrac1{2N}$,  we 
conclude that from an $\eta$-approximation of
$\mbox{S}_N(\B)$, with probability $1-\d$, we can compute the exact value of
$\mbox{S}(\B)$ with probability $1-\d$. We know that, for small~$\eta$,
$A^{{\rm Bool}}(\B,\eta,\d)$ computes an $\eta$-approximation of 
$\mbox{S}_N(\B)$
with probability $1-\d$. Letting  $\eta=1/(3N)$
we compute $A^{{\rm Bool}}_n(\B,1/(3N),\d)$,  which is 
an $1/(3N)$-approximation of $\mbox{S}_N(\B)$. From this we can compute
the exact value of $\mbox{S}_N(\B)$ with probability $1-\d$. This means that
$$
A_n^{{\rm SAT}_1}(\B,\d)\,=\,
\begin{cases}
\mbox{YES}&\text{if}\; \left\lfloor N\,A^{{\rm
Bool}}(\B,1/(3N),\d)+\tfrac12\right\rfloor\,>\,0,\\
\mbox{NO} &\text{if}\; \left\lfloor N\,A^{{\rm
Bool}}(\B,1/(3N),\d)+\tfrac12\right\rfloor\,=\,0
\end{cases}
$$
solves the satisfiability problem. We stress that to compute 
$A_n^{{\rm SAT}}(\B,\d)$ we run the quantum algorithm
$A^{{\rm Bool}}_n(\B,1/(3N),\d)$ and the rest is computed on a
classical computer. Since we know how many power queries and
qubits are used by $A^{{\rm Bool}}_n$ we obtain the following theorem.

\begin{thm}
The satisfiability problem $\mbox{SAT}_1$ for the class $\B_n$ 
is solved with probability $1-\d$ by the quantum algorithm
$A_n^{{\rm SAT}_1}$ which uses of order
\begin{itemize}
\item $n\,\log\d^{-1}$ power queries and
\item $n$ qubits.
\end{itemize}
\end{thm}
\vskip 1pc
We turn to the $\mbox{SAT}_2$ problem. That is, for a non-zero
$\B$ from $\B_n$ we want to compute an index $j\in\{0,1,\dots,N-1\}$ 
for which $\B(j)=1$. 

We will use bisection on the domain of $\B$. Every bisection step will
shrink the cardinality of the domain by $2$.  Using the quantum
algorithm $A_k^{{\rm SAT}_1}$ with $k=n,n-1,\dots,0$, we will know
whether an index of the true assignment 
belongs to the the decreased domain. 
In this way, after $n$
steps we identify an index $j$ for which $\B(j)=1$. Since 
$A_k^{{\rm SAT}_1}$ is a probabilistic algorithm and we use it $n$ times,
we need 
the success probability of this algorithm to be $1-\d_1$, where
$$
(1-\d_1)^n\,=\,1-\d.
$$
For small $\d$, we obviously have $\d_1=\d/n(1+o(1))$. 

More precisely, let
$$
D_k\,=\,\{0,1,\dots,2^k-1\}\qquad\mbox{for}\ \ k=0,1,\dots,n.
$$
We set $j_{n}=0$, and perform the following steps for the Boolean function $\B$
from $\B_n$.
\vskip 1pc
For $k=n-1,n-2,\dots,1,0$ do:
\begin{itemize}
\item define $f_k:D_k\to \{0,1\}$ by $f_k(j)=\B(j+j_{k+1})$ for $j\in D_k$,
\item run the quantum algorithm $A_k^{{\rm SAT}_1}(f_k,\d_1)$ and compute
on a classical computer
$$
j_k\,=\,
\begin{cases}
j_{k+1}+2^k&\text{if}\; A_k^{{\rm SAT}_1}(f_k,\d_1)\,=\,\mbox{NO},\\
j_{k+1} &\text{if}\; A_k^{{\rm SAT}_1}(f_k,\d_1)\,=\,\mbox{YES}.
\end{cases}
$$
\end{itemize}
Finally we set
$$
A_n^{{\rm SAT}_2}(\B,\d)\,=\,j_0.
$$
We claim that the algorithm $A_n^{{\rm SAT}_2}$ solves the  
satisfiability problem $\mbox{SAT}_2$, 
i.e., for the index~$j_0$ we have $\B(j_0)=1$.

Indeed, first   note that
$j_k\,\le\,2^k+2^{k+1}+\cdots+2^{n-1}\,\le\,2^n-1=N-1$ and
$j+j_{k+1}\le 2^k-1+j_{k+1}\le N-1$ for $j\in D_k$.
Therefore, the Boolean functions $f_k$ are well defined. 

For the first step, $k=n-1$, we have $f_{n-1}\equiv \B$ on the first
half, $D_{n-1}$, of the domain $D_n$. 
We check whether $f_{n-1}$ is zero. This holds
with probability $1-\d_1$ iff $A_{n-1}^{{\rm
SAT}_1}(f_{n-1},\d_1)=\mbox{NO}$. If $f_{n-1}\equiv 0$ then $\B$ is
non-zero on the complement of $D_{n-1}$ and an index $j$ for which
$\B(j)=1$ is at least equal to $2^{n-1}$. That is why we define
  $j_{n-1}=2^{n-1}$ in this case.
If, however, $f_{n-1}$ is non-zero over $D_{n-1}$ then 
we are looking for an index j with $\B(j)=1$ in the set $D_{n-1}$,
and we set $j_{n-1}=j_n=0$. 
In this way, after the first step we can restrict the search of an
index to the domain of cardinality $2^{n-1}$.
For the second step it is enough to work with the domain
$D_{n-2}$ and use the proper shift $j_{n-1}$ in the definition
of the Boolean function $f_{n-2}$. 
After $n$ steps we identify a proper index $j=j_0$ for which
$\B(j)=1$. In fact, it is easy to see that we will find the smallest
index $j$ for which $\B(j)=1$ 

Since the quantum algorithm $A_k^{{\rm SAT}_1}
(f_k,\d_1)$ works with probability 
$1-\d_1$ and we repeat $n$ times the algorithm, the probability of
success is at least $(1-\d_1)^n$. By the definition of $\d_1$,
this is equal to $1-\d$. This proves the following theorem.
\begin{thm}
The satisfiability problem $\mbox{SAT}_2$ for the class $\B_n$ is solved
with probability $1-\d$ by the quantum algorithm
$A_n^{{\rm SAT}_2}$ which uses of order
\begin{itemize}
\item
$n^2\,\left(\log\d^{-1}\,+\,\log\,n\right)$ power queries and
\item $n$ qubits.
\end{itemize}
\end{thm}

\section{Grover's problem}

Grover's problem can be defined as the satisfiability problem
$\mbox{SAT}_2$  
for the class $\bar \B_n$ of Boolean functions of $n$ variables
for which we know a priori that there exists exactly one index $j=j_{\B}$ for
which $\B(j)=1$. Obviously, the algorithm $A_n^{{\rm SAT}_2}$
solves Grover's problem although the a priori knowledge
about the uniqueness of the index $j_{\B}$ is not used. 
We now show that using this a priori knowledge it is possible to find a
more efficient quantum algorithm than $A_n^{{\rm SAT}_2}$.

Define the weighted Boolean mean
$$
W_N(\B)\,=\,\frac1{N}\sum_{j=0}^{N-1}j\,\B(j)\qquad\mbox{for}\ \
\B\in\bar \B_n.
$$
Clearly, $W_N(\B)=j_{\B}/N$. Hence, it is enough to compute
the exact value of $W_N(\B)$ and then $j_{\B}=N\,W_N(\B)$. This can be 
achieved by
switching to the integration problem, as we did in Section
\ref{Bms}, for the function $g_{\B}:[0,1]\to\R$ defined by
\begin{equation*}
g_{\B}(x)\,=\,
\begin{cases}
j\,h_j(x)\,\B(j)\,\left(2\,N\,\sin^2(\pi x)\right)^{-1} 
&\text{if}\; x\in[x_j,x_{j+1}] \ \text{with} \ j=0,1,\dots,N-1,\\
0 &\text{if} \; x\in[0,\tfrac14]\cup[\tfrac34,1].
\end{cases}
\end{equation*}
As in Section \ref{Bms}, we conclude that $g_{\B}\in \F_M$ for
$M=\|h^{\prime\prime}\|_{\infty}(1+O(N^{-1}))$, and 
$$
I(g_{\B})\,=\,\frac{j_{\B}}{2N}\int_{x_{j_{\B}}}^{x_{j_{\B}+1}}h_j(x)\,dx\,=\,
\frac{\mbox{Int}(h)}{16N^3}\,W_N(\B).
$$
Hence,
$$
j_{\B}\,=\,\frac{16N^4}{{\rm Int}(h)}\,I(g_{\B}).
$$

In Section \ref{inte}, we defined the algorithm 
$A^{{\rm Int}}$ such that, for small $\e$, $A^{{\rm Int}}(f,\e,\d)$ is an 
$\e$-approximation of $I(f)$ with probability $1-\d$.

Let $\e=\mbox{Int}(h)/(48N^4)$. Define the quantum algorithm 
$$
A_n^{{\rm Grover}}(\B,\d)
\,=\,\left\lfloor
\frac{16N^4}{{\rm Int}(h)}\,
A^{{\rm Int}}(g_{\B},\e,\d)
\,+\,\frac12\right\rfloor.
$$
Then 
$$
\left|j_{\B}-\frac{16N^4}{{\rm Int}(h)}\,
A^{{\rm Int}}(g_{\B},\e,\d)\right|\,=\,
\frac{16N^4}{{\rm Int}(h)}
\left|I(g_{\B})-A^{{\rm Int}}(g_{\B},\e,\d)\right|\,\le\,
\frac{16N^4}{{\rm Int}(h)}\,\e\,=\,\frac13,
$$
and this holds with probability $1-\d$. Hence,
$$
j_{\B}\,=\,A_n^{{\rm Grover}}(\B,\d)\ \
\mbox{with probability}\  1-\d.
$$
This and Theorem \ref{3.1} yield the following theorem.

\begin{thm}
Grover's problem for the class $\B_n$ is solved 
with probability $1-\d$ by the quantum algorithm
$A_n^{{\rm Grover}}$ which uses of order
\begin{itemize}
\item $n\,\log\d^{-1}$ power queries and
\item $n$ qubits.
\end{itemize}
\end{thm}

\section{Minimization}\label{minimization}
In this section we consider a real number minimization problem 
which we will use to solve the traveling salesman problem in the
quantum setting with power queries. 

For positive $N=2^n$ and $M$, define the set
$$
\X_{n,M}\,=\,\{\,x=[x_0,x_1,\dots,x_{N-1}]\,:\ x_j\in\R\ \mbox{and}\
|x_j|\,\le\,M\ \mbox{for}\ \ j=0,1,\dots,N-1\,\}.
$$
Let
$$
\M(x)\,=\,\min_{j=0,1,\dots,N-1}x_j.
$$

The minimization problem is defined as: 
\begin{itemize}
\item\ {\bf $\mbox{MIN}_1:$}\ 
compute $A(x)$ which is an $\e$-approximation of $\M(x)$
with probability $1-\d$, i.e.,  $|\M(x)-A(x)|\,\le\,\e$ holds
with probability $1-\d$ for all $x\in \X_{n,M}$. 
\item\ {\bf $\mbox{MIN}_2:$}\ 
compute an index $j=j(x)$ for which $|\M(x)-x_j|\le \e$ with
probability $1-\d$ for all $x\in \X_{n,M}$.
\end{itemize}

Clearly, using a classical computer we must use each $x_j$ at least
once and that is why the worst case and randomized complexities are
proportional to $N$, i.e., they are exponential in~$n$. We now show
how to solve this problem in the quantum setting using a number
of power queries which is polynomial in 
$n,\log\,M,\log\,\d^{-1}$ and $\log\,\e^{-1}$.

We begin with the minimization problem $\mbox{MIN}_1$.
For a real number $y$, define the Boolean function
$f_y:\{0,1,\dots,N-1\}\to\{0,1\}$ by
$$
f_y(j)\,=\,
\begin{cases}
1 &\text{if}\; x_j\,\le\,y,\\
0 &\text{if}\; x_j\,>\,y.
\end{cases}
$$
Then let
$$
\mbox{S}_N(f_y)\,=\,\frac1N\sum_{j=0}^{N-1}f_y(j).
$$
Note that
$$
\mbox{S}_N(f_y)>0\quad\mbox{iff}\quad y\,\ge\,\M(x).
$$
Clearly, the condition $\mbox{S}_N(f_y)>0$ is equivalent to the $\mbox{SAT}_1$
problem for the Boolean function~$f_y$ and can be solved by 
the quantum algorithm $A^{{\rm SAT}_1}_n$ of Section \ref{sat}. 

Initially, we know that $\M(x)\in[-M,M]$. That is why we set $a_0=-M$,
$b_0=M$ and $y_0=0$, and use the bisection algorithm for the interval
$[-M,M]$ with $k^*$ steps, 
$$
k^*\,=\,\left\lceil \log_2\frac{M}{\e}\right\rceil.
$$
We also choose $\d_1$ such that $(1-\d_1)^{k^*}=1-\d$. For small $\d$,
we have $\d_1=\d/k^*(1+o(1))$. More precisely, we 
perform the following steps.

For $k=1,2,\dots,k^*$ do:
\begin{itemize}
\item run the quantum algorithm $A^{{\rm SAT}_1}(f_{y_{k-1}},\d_1)$,
\item compute on a classical computer
$$
a_k\,=\,
\begin{cases}
a_{k-1} &\text{if}\; A^{{\rm SAT}_1}_n(f_{y_{k-1}},\d_1)=\mbox{YES},\\
y_{k-1} &\text{if}\; A^{{\rm SAT}_1}_n(f_{y_{k-1}},\d_1)=\mbox{NO},
\end{cases}
$$
$$
b_k\,=\,
\begin{cases}
y_{k-1} &\text{if}\; A^{{\rm SAT}_1}_n(f_{y_{k-1}},\d_1)=\mbox{YES},\\
b_{k-1} &\text{if}\; A^{{\rm SAT}_1}_n(f_{y_{k-1}} ,\d_1)=\mbox{NO},
\end{cases}
$$
\item and $y_k\,=\,\tfrac12(a_k+b_k)$.
\end{itemize}
Finally set
$$
A^{{\rm Min}_1}_{n,M}(x,\e,\d)\,=\,y_{k^*}.
$$

After $k^*$ steps we have the interval $[a_{k^*},b_{k^*}]$ of length
$2M/2^{k^*}\le2\e$, and $\M(x)\in[a_{k^*},b_{k^*}]$. Therefore
$|\M(x)-y_{k^*}|\le \e$ is an approximation of $\M(x)$. Note that this
algorithm works with probability $(1-\d_1)^{k^*}=1-\d$. Knowing the
requirements of the algorithm $A^{{\rm SAT}_1}_n$ we obtain 
the following theorem. 

\begin{thm}
We compute an $\e$-approximation with probability 
$1-\delta$ for the minimization problem $\mbox{MIN}_1$ for the class $\X_{n,M}$ 
by the quantum algorithm $A^{{\rm Min}_1}_{n,M}$ using of order
\begin{itemize}
\item
$n\,\left(\log\,M\,+\,\log\,\e^{-1}\right)
\left(\log\d^{-1}\,+\,\log\,\left(\log\,M\,+\,\log\,\e^{-1}\right)\right)$
 power queries and
\item
$n$ qubits.
\end{itemize}
\end{thm}

We now turn to the minimization problem $\mbox{MIN}_2$. It is easy to see that
this problem can be solved by combining the quantum algorithms
developed so far. To explain the main idea of the quantum algorithm
for the minimization problem $\mbox{MIN}_2$
we ignore for a moment the fact that all
the quantum algorithms of the previous sections work probabilistically. 
Knowing $y=y_{k^*}$ by the $A^{{\rm Min}_1}_n$ algorithm, we apply 
the $A^{{\rm Bool}}_n$ algorithm of Section \ref{Bms} and compute 
the exact value of $\mbox{S}_N(f_y)$. If $\mbox{S}_N(f_y)\not=0$ 
then $f_y$ is a non-zero 
Boolean function. If $S(f_y)=0$ then $y<\M(x)$ and since $y$ is 
an $\e$-approximation to $\M(x)$, we have $\M(x)-\e\le y<\M(x)$. 
In this case we have $\M(x)\le y+\e$
and $0\le y+\e-\M(x)\le\e$. Hence, $y+\e$ is also an $\e$-approximation
to $\M(x)$ and $S(f_{y+\e})\not=0$. 
Thus we modify $y:=y+\e$ if $S(f_y)=0$. Then $S(f_y)\not=0$ in either
case, and $y$ is still an $\e$-approximation to $\M(x)$. Knowing that
$f_y$ is a non-zero Boolean function, we now run the $A^{{\rm
SAT}_2}_n$ algorithm and compute an index $j$ for which $f_y(j)=1$,
or equivalently, $x_j\le y$. Hence, $\M(x)\le x_j\le y\le \M(x)+\e$
and $x_j$ is an $\e$-approximation that solves the 
minimization problem $\mbox{MIN}_2$. 
We now formalize the idea of this algorithm.
Let $(1-\d_1)^3=1-\d$. Hence, for small $\d$ we have
$\d_1=3\delta(1+o(1))$. We perform the following steps:
\begin{itemize}
\item run the quantum algorithm $A^{{\rm Min}_1}_{n,M}(x,\e,\d_1)$ to obtain
$y$,
\item run the quantum algorithm $A^{{\rm Bool}}_n(f_y,1/(3N),\d_1)$ to obtain
$z$,
\item if $z=0$ then set $y:=y+\e$,
\item run the quantum algorithm $A_n^{{\rm SAT}_2}(f_y,\d_1)$ to
obtain $j$.
\end{itemize}
Finally set,
$$
A_{n,M}^{{\rm Min}_2}(x,\e,\d)\,=\,j.
$$
Then the index $j$ solves the minimization problem $\mbox{MIN}_2$ and this
algorithm works with probability $(1-\d_1)^3=1-\d$. Counting the
number of power queries and qubits of all parts of the algorithm we obtain
the theorem.

\begin{thm}
We compute an $\e$-approximation with probability 
$1-\delta$ for the minimization problem $\mbox{MIN}_2$ 
for the class $\X_{n,M}$ by the quantum algorithm $A^{{\rm Min}_2}_{n,M}$ 
using of order
\begin{itemize}
\item 
$n^2\left(\log\,\d^{-1}+\log\,n\right)+ 
n\,\left(\log\,M\,+\,\log\,\e^{-1}\right)
\left(\log\d^{-1}\,+\,\log\,\left(\log\,M\,+\,\log\,\e^{-1}\right)\right)
$ \linebreak
power queries and
\item
 $n$ qubits
\end{itemize}.
\end{thm}

\section{Traveling salesman}\label{TSP}
 
The traveling salesman problem, TSP for short, is a well known
NP-complete problem that deals with the shortest tour between $m$
cities, see \cite{garey}. 
Let $D=[d(j,k)]_{j,k=1}^m$ denote the $m\times m$ matrix with
$d(j,k)$ being the distance between the city $j$ and the city $k$. We
assume that $d(j,k)$ are positive integers for $j\not= k$ and $d(j,j)=0$. 
The TSP can be studied as a decision or as a computational problem
and we consider the three variants:
\begin{itemize}
\item\ {\bf $\mbox{TSP}_1:$}\ for a given integer $B$ verify if there is a
permutation $\pi=[\pi(1),\pi(2),\dots,\pi(m)]$ of indices
$\{1,2,\dots,m\}$ for which
$$
d(\pi)\,:=\,\sum_{j=1}^{m}d(\pi(j),\pi(j+1))\,\le\,B
$$
with $\pi(m+1)=\pi(1)$. 
\item\ {\bf $\mbox{TSP}_2:$}\ compute
$$
\M(D)\,=\,\min_{\pi}d(\pi).
$$
\item\ {\bf $\mbox{TSP}_3:$}\ compute a permutation $\pi^*$ such that
$$
d(\pi^*)\,=\,\M(D).
$$
\end{itemize}
Observe that if we can solve $\mbox{TSP}_2$ then it is just enough to
check whether $\M(D)\le B$. Similarly, if we can solve $\mbox{TSP}_3$
then it is enough to compute $d(\pi^*)$ to solve $\mbox{TSP}_2$.  

We now show that TSP is a special case of the minimization problem
studied in the previous section. Indeed, consider the set $P_m$ of
all $m!$ possible permutations, and let $g:\,\{0,1,\dots,m!-1\}\to P_m$ be
an injective mapping. We take 
$$
n\,:=\,\lceil \log\,m!\rceil\,=\,m\log\,m\, (1+o(1)).
$$
Let $N=2^n$. For $j=m!,m!+1,\dots,N-1$ we extend the function $g$ by setting 
$g(j)=g(m!-1)$. Then
$g:\{0,1,\dots,N-1\}\to P_m$ and 
$$
g(j)\,=\,\pi_j\,=\,[\pi_j(1),\pi_j(2),\dots,\pi_j(m)]\qquad\mbox{for}\
\ j\in\{0,1,\dots,N-1\}.
$$
Defining
$$
x_j\,=\,\sum_{k=1}^md(\pi_j(k),\pi_j(k+1))\qquad\mbox{with}\ \
\pi_j(m+1)\,=\,\pi_j(1),
$$
we see that 
$$
\min_{j=0,1,\dots,N-1}x_j\,=\,\M(D).
$$
Note that $x_j$ can be computed using a number of bits which is
polynomial in $m$ and the maximal distance between cities, 
$$
d_{\max}\,=\,\max_{i,j=1,2,\dots,m}d(i,j).
$$
Furthermore, $x_j\in [0,M]$ for $j\in \{0,1,\dots,N-1\}$ with  
$$
M\,\ge\,m\,d_{\max}.
$$
To run the algorithms $A_{n,M}^{{\rm Min}_1}$ and $A_{n,M}^{{\rm Min}_2}$
for the solutions of the minimization problem we need to know
an upper bound on $m\,d_{\max}$. This can be achieved as follows. 
For $k=0,1,\dots$, we run the quantum algorithm
$A_n^{{\rm SAT}_1}(1-f_{2^k},\delta_1)$ of Section \ref{sat}
to check, with probability $1-\delta_1$, whether
the Boolean function $1-f_{2^k}$ defined in Section \ref{minimization}
is zero.  If so, then $f_{2^k}\equiv 1$ meaning that for all $x_j\le
2^k$, and we can take $M=2^k$. Hence, we perform as  many steps 
as necessary until
the first occurrence of $A_n^{{\rm SAT}_1}(1-f_{2^k},\delta_1)=\mbox{NO}$.
This will happen after at most $p:=\lceil
\log_2\,m+\log_2\,d_{\max}\rceil$ steps. Then we define $M=2^p$ and
$M\in[m\,d_{\max},2m\, d_{\max}]$.  
We choose $\d_1$ such that $(1-\d_1)^p=(1-\d)^{1/2}$ 
to conclude that we obtain $M$ with probability
$(1-\d)^{1/2}$. For small $\d$, we have $\d_1=\d/(2p)(1+o(1))$. 

Knowing $M$, we can solve TSP by solving the minimization problem of Section
\ref{minimization}.  Since all $x_j$ are integers, 
it is enough to compute an $\e$-approximation $y$ of $\M(x)$ with
$\e<\tfrac12$ to conclude that $\lfloor y+\tfrac12\rfloor$ is equal to
$\M(x)$. The number $y$ can be computed with probability $(1-\d)^{1/2}=1-\d_2$
by the quantum algorithm $A^{{\rm Min}_1}_{n,M}(x,\e,\d_2)$ with $\e$, say,
$\tfrac13$. In the same way, we can find a shortest path represented by
a permutation $\pi^*$ by first computing the index~$j$ by the algorithm
$A^{{\rm Min}_2}_{n,M}$ of Section \ref{minimization} for which
$|\M(x)-x_j|\le\tfrac13$. Again, since $x_j$ is an integer,
$x_j=\M(x)$ and $\pi^*=g(j)$ is a needed permutation. Hence
the following algorithms
\begin{eqnarray*}
A^{{\rm TSP}_1}_m(D,\delta)\,&=&\,
\begin{cases}
\mbox{YES} &\text{if}\; \lfloor A^{{\rm
Min}_1}_{n,M}(x,\tfrac13,\d_2)+\tfrac12\rfloor
\,\le\,B,\\
\mbox{NO}  &\text{if}\; \text{otherwise},
\end{cases}\\
A^{{\rm TSP}_2}_m(D,\delta)\,&=&\,x_{A^{{\rm Min}_2}_{n,M}(x,\tfrac13,\d_2)}
\,=\,d(\pi_{A^{{\rm Min}_2}_{n,M}(x,\tfrac13,\d_2)})
\\
A^{{\rm TSP}_3}_m(D,\delta)\,&=&\,g\left(A_{n,M}^{{\rm
Min}_2}(x,\tfrac13,\d_2)\right)
\end{eqnarray*}
solve the traveling salesman problems $\mbox{TSP}_j$ for $j=1,2,3$,
respectively, with probability $1-\d_2=(1-\d)^{1/2}$. 
We also know the proper $M$ with probability $(1-\d)^{1/2}$.
Therefore, the whole algorithm works with probability $1-\d$.
This proves the following theorem. 

\begin{thm}
The traveling salesman problems $\mbox{TSP}_j$ are solved  with
probability $1-\d$ 
by the quantum algorithms $A^{{\rm TSP}_j}_m$ using of order 
\begin{itemize}
\item for $j=1$,
  \begin{itemize}
   \item
 $m\log\,m\left(\log\,\d^{-1}\,+\,\log\,m\,+\,\log\,d_{\max}\right)
(\log\,m+\log\,d_{\max})$  power queries and
   \item $m\log\,m$ qubits,
\end{itemize} 
\item for $j=2,3$, 
\begin{itemize}
\item 
\begin{eqnarray*}
m^2\log^2&m&\left(\log\,\d^{-1}+\log\,m\,\right)\\
&+&\,
m\log\,m\left(\log\,\d^{-1}+\log\,m+\log\,d_{\max}\right)
(\log\,m+\log\,d_{\max})
\end{eqnarray*}
 power queries and 
\item $m\log\,m$ qubits.
\end{itemize}
\end{itemize}
\end{thm}
 
\section*{Acknowledgments}

This research has been supported in part by 
the National Science Foundation, the Defense Advanced Research
Projects Agency, and the Air Force  
Research Laboratory. 

We are grateful for valuable comments and suggestions how to improve
the presentation of the paper from Stefan Heinrich, Marek Kwas, Klaus Meer
Joseph F. Traub and Mihalis Yannakakis.

\end{document}